%% file: main.tex
\documentclass{article}
\usepackage{arxiv}

\usepackage[utf8]{inputenc} % allow utf-8 input
\usepackage[T1]{fontenc}    % use 8-bit T1 fonts
\usepackage{hyperref}       % hyperlinks
\usepackage{url}            % simple URL typesetting
\usepackage{booktabs}       % professional-quality tables
\usepackage{amsfonts}       % blackboard math symbols
\usepackage{nicefrac}       % compact symbols for 1/2, etc.
\usepackage{microtype}      % microtypography
\usepackage{lipsum}		% Can be removed after putting your text content
\usepackage{graphicx}
\usepackage{natbib}
\usepackage{doi}
\usepackage{tikz}
\usepackage{forest}
\usepackage{amsmath}
\usepackage{amsthm}
\usepackage{amssymb}
\usepackage{cancel}
\usepackage{algorithm}
\usepackage{algpseudocode}
\usepackage{multirow}
\usepackage{comment}
\newtheorem{theorem}{Theorem}
\newtheorem{claim}{Claim}

\usepackage{placeins} % help with formatting experiments before bib

\title{Fast Rational Search via Stern–Brocot Tree}

\author{Connor Weyers\thanks{Work done while at UNL School of Computing and supported in part by a UNL Grand Challenges Catalyst Competition Grant} \\
	Department of Computer Science\\
	Indiana University\\
	\texttt{coweyers@iu.edu} \\
	\And
	\href{https://cse.unl.edu/~vinod/}
    {N. V. Vinodchandran\thanks{Supported in part by NSF grants 2342244, 2413848
and a UNL Grand Challenges Catalyst Competition Grant.}} \\
	School of Computing\\
	University of Nebraska-Lincoln\\
	\texttt{vinod@unl.edu} \\
}

\hypersetup{
pdftitle={A template for the arxiv style},
pdfsubject={q-bio.NC, q-bio.QM},
pdfauthor={David S.~Hippocampus, Elias D.~Striatum},
pdfkeywords={First keyword, Second keyword, More},
}

\begin{document}

\maketitle
\begin{abstract}
%We revisit the problem of rational search: given an unknown rational number $\alpha = {a\over b} \in [0,1]$ where $b\leq n$, find $\alpha$ by making queries of the form ``$\beta\leq \alpha$''. We propose a new algorithm  based on a compressed Stern-Brocot tree search which appears to have overlooked in the literature. The new algorithm also solves the problems of unbounded rational search where $n$ is not known and finding the {\em best} (in certain sense) rational approximation of an {\em unknown} real $\alpha$. These problems have not been treated in the literature.   

We revisit the problem of rational search: given an unknown rational number \(\alpha = \frac{a}{b} \in (0,1)\) with \(b \leq n\), 
the goal is to identify \(\alpha\) using comparison queries of the form ``\(\beta \leq \alpha\)?''. The problem has been studied several decades ago and optimal query algorithms are known. We present a new algorithm for rational search based on a compressed traversal of the Stern--Brocot tree, which appeared to have been overlooked in the literature.  This approach also naturally extends to two related problems that, to the best of our knowledge, have not been previously addressed: 
(i) unbounded rational search, where the bound \(n\) is unknown, and 
(ii) computing the \emph{best} (in a precise sense) rational approximation of an \emph{unknown} real number using only comparison queries. 
\end{abstract}

\input{intro}
\input{SBTree}
%\input{farey}
\input{algorithm}

\input{experiments-writeup}
\input{experiments}

%\newpage

%\newpage
%\newpage
\FloatBarrier %maybe \clearpage instead
\bibliographystyle{alpha}
\bibliography{references}
\end{document}

%% file: intro.tex
\section{Introduction}

Rational search is the problem of identifying an unknown fraction from a set of fractions. 
In the \emph{bounded} version of the problem, the input is an unknown rational number 
\(\alpha = \frac{a}{b} \in [0,1]\) with \(b \leq n\), and the goal is to determine \(\alpha\) using 
comparison queries of the form ``\(\beta \leq \alpha\)?''. 
The \emph{unbounded} version assumes no knowledge of \(n\): $\alpha$ could be any rational number in $[0,1]$.  The efficiency of a rational search algorithm is typically measured by the number of such queries needed to uniquely identify the target fraction.

This problem was first studied several decades ago, and algorithms for the bounded version were proposed 
in~\cite{PAPADIMITRIOU19791,reiss1979rational}, culminating in an query-optimal (up to an additive constant) algorithm 
by Kwek and Mehlhorn in 2003~\cite{kwek2003optimal}. 
The Kwek--Mehlhorn (KM) algorithm requires at most \(2\log_2 n + O(1)\) queries, which is optimal up to an additive constant, 
since the number of distinct fractions in \([0,1]\) with denominator \(\leq n\) is \(\Omega(n^2)\); 
see, for example,~\cite{graham1994concrete}.  

In this note, we present an algorithm for the rational search problem. To the best of our knowledge this approach to rational search is not treated in the literature. 
The algorithm is intuitive, based on a compressed traversal of the rational data structure known as the 
\emph{Stern--Brocot tree}. 
While simple to describe and easy to implement, its analysis is somewhat non-trivial. 
We prove an upper bound of \(2.5849\log_2 n\) comparisons in the worst case 
(where \(2.5849\) is an approximation to the exact constant \(16/\log_2 73\) arising in the analysis). 
We also construct instances that require \(2.4189\log_2 n\) comparisons 
(where \(2.4189\) approximates the irrational \(8/\log_2(5 + 2\sqrt{6})\)). Finding the exact worst-case complexity of the algorithm is open. 
Based on experiments we conducted, we conjuncture that the correct upper bound is that given by our lower bound.  

The algorithm is not worst-case query-optimal for the bounded rational search. 
However, an advantage is that our approach naturally solves the most general unbounded rational search problem. We note that while we focus on unbounded search of rationals in $(0,1)$, the same algorithm works for searching for any unknown positive rational number in $\mathbb{Q}$ using comparison queries without any additional knowledge. 
Moreover, it can be easily adapted to solve the problem of finding the \emph{best}  rational approximation of an 
\emph{unknown} real number using only comparison queries, discussed next. To the best of our knowledge, neither of these two variations has been previously addressed in the literature.

%We present an algorithm that addresses both the rational search problem and the rational approximation problem for an \emph{unknown} real number. It uses a compressed traversal of the rational data structure known as the Stern-Brocot tree.

A problem related to the rational search problem is that of finding the \emph{best} rational approximation to a given real (or irrational) number.
Specifically, given a real number $r$ and an approximation parameter $\delta$, find the (unique) fraction ${a\over b}$ so that ${a \over b} \in [r\pm \delta]$ with the \emph{smallest possible denominator} $b$. This problem is non-trivial even  when $r$ is given explicitly. In fact, Fori\v{s}ek proposed an efficient algorithm for this problem, when $r$ is given, with the number of arithmetic operations linear in the input representation~\cite{forivsek2007approximating}. To the best of our knowledge, no efficient query-based algorithm is known for the case when the target real number is \emph{unknown} and must be identified solely through comparison queries. An easy modification of our algorithm also solves this best rational approximation problem for an \emph{unknown} real number. While our algorithm is similar to Fori\v{s}ek's approach, it relies on a combination of exponential search and binary search to navigate the Stern-Brocot tree efficiently only with comparison queries.

%Additionally, we conduct an experimental evaluation comparing our algorithm with the Kwek–Mehlhorn algorithm. The results indicate that, while not substantial, our algorithm takes fewer queries on average \footnote{We allow the algorithm to return the fraction as soon as the comparison returns equality.} for values of $n$ up to $10^{40}$.  We also implemented the rational approximation variant of the algorithm and report approximations of the real numbers \( \pi, e, \sqrt{2} \), and \( \sqrt{5} \) (treated as unknown and only using comparison queries) for various values of the approximation parameter, along with the corresponding query counts and runtimes.

We conducted an experimental evaluation comparing our algorithm with the Kwek--Mehlhorn algorithm. 
The results show that, although the improvement is not large, our algorithm requires fewer queries on average\footnote{We allow the algorithm to terminate as soon as a comparison returns equality.} 
when given a denominator's upper bound \( n \) up to \( 10^{40} \). 
We also implemented the rational-approximation variant of our algorithm and evaluated it on the real numbers \( \pi, e, \sqrt{2} \), and \( \sqrt{5} \), 
treating each as unknown and using only comparison queries. 
For various settings of the approximation parameter, we report the resulting approximations along with the corresponding query counts and runtimes.

%% file: SBTree.tex
\section{Stern--Brocot Tree}
The {\em Stern--Brocot tree}, introduced independently by Stern \cite{stern1858} and Brocot \cite{brocot1860} in the 19th century, is an infinite binary tree-structure that systematically enumerates all positive rational numbers in reduced form, each exactly once. The tree is constructed using the \emph{mediant} operation, a simple binary operation on fractions. Given two fractions \( \frac{a}{b} \) and \( \frac{c}{d} \), the {\em mediant} is defined as:
$\frac{a + c}{b + d}.$ This new fraction lies strictly between \( \frac{a}{b} \) and \( \frac{c}{d} \). In general, this fraction is not necessarily in the reduced form, but in the Stern--Brocot tree construction, all mediants turn out to be in the reduced form.  

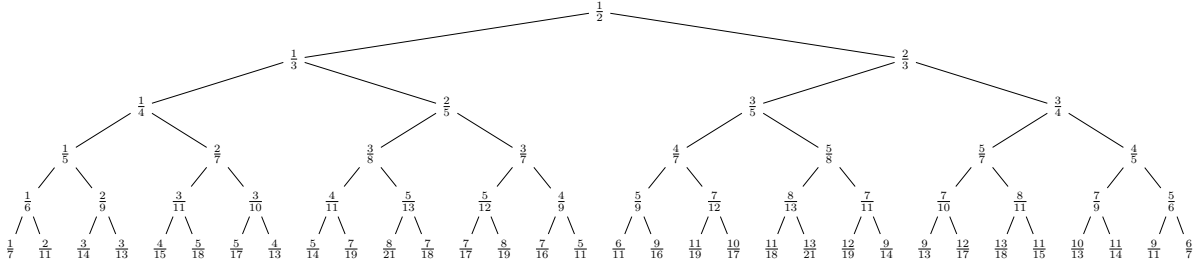
\begin{figure}[h]
\begin{center}
\scalebox{.6}{
\begin{forest}
  Stern Brocot/.style n args={5}{%
    content=$\frac{\number\numexpr#1+#3\relax}{\number\numexpr#2+#4\relax}$,
    if={#5>0}{% true
      append={[,Stern Brocot={#1}{#2}{#1+#3}{#2+#4}{#5-1}]},
      append={[,Stern Brocot={#1+#3}{#2+#4}{#3}{#4}{#5-1}]}
    }{}}% false (empty)
[,Stern Brocot={0}{1}{1}{1}{5}]
\end{forest}}

\caption{\centering Stern-Brocot Tree up to level 6, starting at 1/2. As examples ${1\over 3}={0+1\over 1+2}$ is the mediant of ${0\over 1}$ (not shown) and ${1\over 2}$, and ${5\over 7}={2+3\over 3+4}$ is the mediant of ${2\over 3}$ and ${3\over 4}$. }
\label{fig:sb}
\end{center}
\end{figure}

The tree is initialized with two \emph{sentinel} fractions: $\frac{0}{1} \text{~and~}  \frac{1}{1}$ (general construction of rational starts with $\frac{0}{1} \text{~and~} \frac{1}{0}=\infty$, but we will focus on fractions in \((0,1)\)). The root node is $1/2$ which is the mediant of $\frac{0}{1} \text{~and~}  \frac{1}{1}$. For a leaf $\frac{c}{d}$ of the partial tree constructed so far, let $\frac{a}{b}$ and $\frac{e}{f}$ be elements `left' and `right' (with respect to in-order traversal of the tree). Then left child of $\frac{c}{d}$ is $\frac{a+c}{b+d}$, the mediant of $\frac{a}{b}$ and $\frac{c}{d}$. The right child of 
$\frac{c}{d}$ is $\frac{c+e}{d+f}$, the mediant of $\frac{c}{d}$ and $\frac{e}{f}$. Figure~\ref{fig:sb} gives an illustration up to level 6. 
To build finite Stern--Brocot tree containing only fractions with denominators up to  \( n \), the construction 
halts at any pair \( \frac{a}{b}, \frac{c}{d} \) for which the mediant \( \frac{a + c}{b + d} \) satisfies \( b + d > n \). 

The Stern-Brocot tree contains all positive reduced rational numbers exactly once. Because of this any rational can be uniquely represented by the unique path from the root to the rational number, for example as a sequence of $L$ (left) and $R$ (right).  It is also well known that  Stern-Brocot tree is tightly connected to {\em continued fractions}. Every path in the compressed from the root of the Stern-Brocot tree to a rational number corresponds to its continued fraction representation, and vice versa. This is discussed further below in subsection~\ref{subsec:cf}. The literature on the Stern–Brocot tree and its connections to various number-theoretic concepts is both classical and extensive.
An excellent starting point is the book {\em Concrete Mathematics} by Graham, Knuth, and Patashnik \cite{graham1994concrete}.

We are interested in the binary-search-tree property of the Stern-Brocot tree. Indeed, this structure provides a binary search mechanism for locating any unknown rational number. However, a naive search of walking down the tree one step at a time will result in $n-1$ comparison queries in the worst case (for example to search for $1/n$) which is far from optimal. We present a compressed search algorithm that leads to an algorithm that is much more efficient.

%\begin{figure}[h]
%\begin{center}
%\begin{tikzpicture}[scale=1, every node/.style={font=\footnotesize}]

% Level 0
%\node (a) at (-1,0) {$\frac{0}{1}$};
%\node (b) at (9,0) {$\frac{1}{1}$};

% Level 1
%\node (c) at (4,-1.5) {$\frac{1}{2}$};

% Level 2
%\node (d) at (2,-3) {$\frac{1}{3}$};
%\node (e) at (6,-3) {$\frac{2}{3}$};

% Level 3
%\node (f) at (1,-4.5) {$\frac{1}{4}$};
%\node (g) at (3,-4.5) {$\frac{2}{5}$};
%\node (h) at (5,-4.5) {$\frac{3}{5}$};
%\node (i) at (7,-4.5) {$\frac{3}{4}$};

% Level 4
%\node (j) at (0,-6) {$\frac{1}{5}$};
%\node (k) at (4,-6) {$\frac{3}{4}$}; % Already added above
%\node (l) at (8,-6) {$\frac{4}{5}$};

% Edges
%\draw (a) -- (c);
%\draw (c) -- (b);
%\draw[dotted] (a) -- (d);
%\draw (d) -- (c);
%\draw (c) -- (e);
%\draw[dotted] (e) -- (b);

%\draw (d) -- (f);
%\draw (d) -- (g);
%\draw (e) -- (h);
%\draw (e) -- (i);

%\draw (f) -- (j);
%\draw (i) -- (l);
%\draw[dotted] (i) -- (b);

%\end{tikzpicture}

%\end{center}

%\end{figure}

%The tree can be constructed as follows: Starting from the pair \( \left(\frac{0}{1}, \frac{1}{0}\right) \), perform the following recursively:
%\begin{itemize}
%  \item[-] Compute the mediant \( \frac{a + c}{b + d} \) of each adjacent pair \( \frac{a}{b}, \frac{c}{d} \).
%  \item[-] Insert the mediant between them.
%  \item[-] Recursively apply the process to the two new intervals:
%  \(
%  \left(\frac{a}{b}, \frac{a + c}{b + d}\right) \text{~and~} \left(\frac{a + c}{b + d}, \frac{c}{d}\right).
%  \)
%\end{itemize}

%% file: algorithm.tex
\section{Compressed Stern-Brocot Tree Search Algorithm}

Our algorithm traverses the Stern-Brocot tree in a compressed manner to search for the unknown fraction $\alpha$. As mentioned, any fraction $\alpha \in (0,1)$ can be uniquely represented as a sequence of `left' $(L)$ and `right' $(R)$ traversal of the Stern-Brocot tree starting at 1 (hence the first element of the sequence is $L$). This can be written in the exponential (or compressed) form as a string of alternating traversals: 
\(L^{x_1}R^{x_2}L^{x_3}R^{x_4}\cdots D^{x_l}\) where $D\in \{L,R\}$. For example, ${9/14}$ is represented by $LRLR^3$ (Figure~\ref{fig:sb}). 
If \( \alpha\) is known to be between two fractions \( \frac{p}{q} \) and \( \frac{p'}{q'} \), a naive approach would take the mediant  \( \frac{p+p'}{q+q'} \) each step down and compare to \(\alpha\). However, this approach will result in $\Omega(n)$ queries in the worst case, as mentioned earlier. This is because, to reach the fraction corresponding to $L^x$, we are making $x$ queries using this simple approach. Instead, we can do an {\em unbounded binary search} using comparisons to get to $L^x$. If we go to the `left' $x$ steps by using {\em repeated mediant}, the new fraction to compare will be \(\frac{p'+xp}{q'+xq}\). For the  unbounded binary search, Bentley and Yao gives sophisticated options~\cite{bentley1976almost}. However, our algorithm employs the simplest and well known of them, perform an exponential search to find a bound and then do a binary within the bound. 
%well known approach of first  exponentially increasing the search space and then a binary search to compute the fraction after $D^{x_i}$ for every $i$. 
Suppose we computed the fraction corresponding to 
\(L^{x_1}R^{x_2}\cdots D^{x_{i-1}}\).
To compute the next fraction \(L^{x_1}R^{x_2}\cdots D^{x_{i}}\), we must find the integer $x_i$. The exponential search involves making queries at 
\(L^{x_1}R^{x_2}\cdots D^{2^j-1}\) for each $j\in[1,k+1]$ such that $2^k - 1 < x_i\leq 2^{k+1}-1$ where $k=\lfloor \log_2 x_i\rfloor$. At each iteration the fraction queried is $2^j-1$ repeated mediant in the direction of traversal. Algorithm ~\ref{alg:exponential_search} describes this exponential search subprocedure. The exponential search requires a maximum of $k+1$ queries with $k=\lfloor \log_2x_{i} \rfloor$.
After we find $k$ so that $x_i\in(2^k,2^{k+1}]$, we do a binary search to find the  $x_i$ within this range. We stop the binary search when the $\alpha$ is located between the fraction at $x_i$ and $x_i-1$ steps. This binary search will take a maximum of $\lfloor \log_2x_{i} \rfloor$ queries. Thus the exponential and binary search steps combined will take $2\lfloor \log_2x_i \rfloor + 1$ queries.
If at any point, the queried fraction is equal to $\alpha$, the search stops. Algorithm ~\ref{alg:rational_search} describes this procedure.
It calls Algorithm~\ref{alg:exponential_search} as a subroutine. As written, it is an unbounded rational search algorithm where the bound on the denominator of the rational $\alpha$ we are searching for is not given. For searching $\alpha$, the algorithm is initially called with parameters $(\alpha,{0\over 1}, {1\over 1}, {\rm LEFT})$ where $\alpha$ is accessed using comparison queries.

%\subsection{The Algorithm}

\begin{algorithm}[h!]
\caption{Exponential Search}
\label{alg:exponential_search}
\begin{algorithmic}[1]
\Require Comparisons to Rational $\alpha$
\Require Rational $\frac{p}{q}$, Rational $\frac{p'}{q'}$, Direction $D \in \{\mbox{LEFT,RIGHT}\}$
\Ensure $\frac{p}{q} < \frac{p'}{q'}$
\Procedure{\text{exponential\_search}}{$n,\alpha,\frac{p}{q},\frac{p'}{q'},D$} %Rational $\alpha$,  Rational $\frac{p}{q}$, Rational $\frac{p'}{q'}$, Direction)
    \State $i \gets 1$
    \If{$D$ is \textbf{RIGHT}}
        \While{$\frac{p+(2^i-1)p'}{q+(2^i-1)q'}< \alpha$}
            \State $i \gets i + 1$
        \EndWhile
    \EndIf
    \If{$D$ is \textbf{LEFT}}
        \While{$\frac{p'+(2^i-1)p}{q'+(2^i-1)q} > \alpha$}
            \State $i \gets i + 1$
        \EndWhile
    \EndIf
    \State \Return $i$
\EndProcedure
\end{algorithmic}
\end{algorithm}
 
\begin{algorithm}[h!]
\caption{Compressed Stern-Brocot Tree Search}
\label{alg:rational_search}
\begin{algorithmic}[1]
\Require Comparisons to  Rational $\alpha$
\Require Rational $\frac{p}{q}$, Rational $\frac{p'}{q'}$, Direction $D \in \{\mbox{LEFT,RIGHT}\}$
\Ensure $\frac{p}{q} < \frac{p'}{q'}$
\Procedure{\text{rational\_search}}{$\alpha,\frac{p}{q},\frac{p'}{q'},D$} %Rational $\alpha$,  Rational $\frac{p}{q}$, Rational $\frac{p'}{q'}$, Direction)
\State $k \gets \text{\sc exponential\_search}(\alpha, \frac{p}{q}, \frac{p'}{q'}, D)$
\State $x \gets \text{\sc binary\_search}(\alpha, \frac{p}{q}, \frac{p'}{q'}, 2^{k-1}-1, 2^{k}-1, D)$
\If {$D$ is LEFT}
    \If {$\alpha = \frac{p'+xp}{q'+xq}$}
        \State \Return $\frac{p'+xp}{q'+xq}$
    \EndIf
    \State $result \gets \text{\sc rational\_search}(\alpha, \frac{p'+xp}{q'+xq}, \frac{p'+(x-1)p}{q'+(x-1)q}, \text{RIGHT})$
\Else
    \If {$\alpha = \frac{p+xp'}{q+xq'}$}
        \State \Return $\frac{p+xp'}{q+xq'}$
    \EndIf
    \State $result \gets \text{\sc rational\_search}(\alpha, \frac{p+(x-1)p'}{q+(x-1)p'}, \frac{p+xp'}{q+xq'}, \text{LEFT})$
\EndIf
\State \Return $result$
\EndProcedure
\end{algorithmic}
\end{algorithm}

\subsection{Connection to Continued Fractions}
\label{subsec:cf}
The compressed representation of a rational number has close connection to its {\em continued fraction representation}. Again, Graham, Knuth, and Patashnik [GKP94] gives an excellent treatment of the Stern-Brocot tree representation and the continued fraction representation with points to its history. Irwin~\cite{irwin1989geometry} gives a geometric view of the continued fractions. 

A \emph{continued fraction} is an expression of the form
$
\left( a_0 + 1/\left( a_1 + 1/\left(a_2+ 1/\left ( a_3+ 1/\left(\cdots\right )\right)\right)\right)
\right )$
%\[
%a_0 + \cfrac{1}{\,a_1 + \cfrac{1}{\,a_2 + %\cfrac{1}{\ddots}}},
%\]
where $a_0 \in \mathbb{Z}$ and $a_i \in \mathbb{Z}_{>0}$ for all $i \ge 1$.  
It is customary to denote this expression in the compact form
$[a_0; a_1, a_2, a_3,\ldots ]$. 
Every real number admits a unique continued fraction expansion which is finite if and only if the number is rational.  For example, $5/12$ is represented by $[0;2,2,2]$ and $9/14$ is represented by $[0;1,1,1,4]$ ($a_0$ is 0 for fractions in $[0,1)$).  
For a (known) rational number $p/q$, its continued fraction representation can be efficiently computed by invoking the Euclid's algorithm.

For a fraction with continued fraction expansion $[0; a_1, a_2, \ldots, a_k]$, its compressed representation is
$L^{a_1} R^{a_2} L^{a_3} \cdots D^{a_k-1}$
truncated appropriately so that the final block uses $a_k - 1$ repetitions. For example for $9/14$ the compressed Stern-Brocot tree representation can be read out directly from its continued fraction representation as $L^1R^1L^1R^3$ (the first $L$ being following `left' from $1\over 1$) which matches what we observe from Figure~\ref{fig:sb}.  

Thus, if the rational number is known explicitly, the Euclid's algorithm immediately yields this compressed representation. In contrast, our goal is to compute the representation using only comparison queries to the unknown fraction. In this setting, the problem becomes information-theoretic: we must iteratively reduce the set of possible values of the fraction using only comparison queries. We give such an algorithm and analyze the number of quires needed to compute the representation.

\subsection{A ${2.5849\log_2 n}$ Upper Bound}

While the algorithm is simple and intuitive, obtaining an accurate bound on the worst-case number number of comparisons appears more complicated. We show a bound of ${2.5849\log_2 n}$. The constant 2.5849 is an approximation to the irrational number $16/(\log_2 73)$. We also show instances where $2.4189\log_2 n$ comparisons are required. We will do the analysis for the unbounded version. Analysis of the bounded version will result in a slightly lower number of comparisons since the algorithm would not search past the bound.

\begin{theorem}
\label{thrm:rat_search_bound}
    Let $\alpha$ be an unknown rational number in $(0,1)$. The algorithm {\rm Compressed Stern-Brocot Tree Search} outputs $\alpha$  with no more than \(2.5849\log_2n \) queries of the form  \(``\beta \leq \alpha"\) where $n$ is the denominator of $\alpha$.
\end{theorem}

First we introduce some notation and state some properties.  The {\em unknown} rational \( \alpha \) can be represented as a sequence \(L^{x_1}R^{x_2}L^{x_3}R^{x_4}\ldots D^{x_l}\) for some $j$ where $D \in \{L,R\}$ is the `direction' of the last {\em segment}. For an $i:1\leq i\leq l$, let 
    \( d_i \) to be the denominator of the fraction represented by \(L^{x_1}R^{x_2} \ldots D^{x_i}\). 
   Let \(m_i\)  be the denominator of the fraction at one less traversal than \(d_i \): that of \(L^{x_1}R^{x_2} \ldots D^{x_i-1}\). Define $d_0=1$ and $m_0 =1$.
   We write some crucial properties as the following claims:
    \begin{claim}\label{claim:d}
        (1) \( d_0=1, d_l \leq n \) and \( d_i = d_{i-1} + x_im_{i-1} \) 
        (2) \(m_0 = 1\) and \( m_i = d_{i-1} + (x_i - 1)m_{i-1} \). From this, it follows that $d_i = m_i + m_{i-1}$ and $m_i/d_i\geq 1/2$. 
    \end{claim}

\begin{claim}\label{claim:query}
For any $i$, given $d_{i-1}$ and $m_{i-1}$, the algorithm computes $d_i$ and $m_{i}$ (and $x_i$) using at most $2\lfloor\log_2 x_{i}\rfloor + 1$  queries. 
\end{claim}

For simplicity, we denote the function $2\lfloor\log_2 x\rfloor + 1$ as $G(x)$. 

\paragraph*{Warm up: $3.1547\log_2 n$ Upper Bound}

We first show a simpler analysis that results in a weaker bound. Fix an arbitrary sequence $x_1,\ldots,x_l$ representing a rational as described before. We will show, for an appropriate constant $C$ the following holds. We will choose $C$ later but it will be greater than 3. 
\[
\sum_{i=1}^l G(x_i) \leq C\log_2 d_l
\]
Since $d_l\leq n$, the bound follows. We will prove this using induction on $l$. For $l=1$,  $2\lfloor\log_2 x_1\rfloor + 1 \leq C\log_2(x_1+1)$ for all $x_1\geq 1$. Assume the hypothesis is true for $l-1$. Then we have to show 
\[
\sum_{i=1}^{l-1} G(x_i) + 2\lfloor\log_2 x_l\rfloor + 1 \leq C\log_2 d_l. 
\]
Using the hypothesis, it is sufficient to show: 
\[
C\log_2 d_{l-1} + 2\lfloor\log_2 x_l\rfloor + 1 \leq C\log_2 d_l. 
\]

\noindent That is, we need a $C$ so that 
\[
C\log_2 \left({d_l\over d_{l-1}}\right) = C\log_2 \left({d_{l-1} +x_lm_{l-1}\over d_{l-1}}\right) \geq  2\lfloor\log_2 x_l\rfloor + 1 .
\]
Using the fact that ${m_{l-1}\over{d_{l-1}}} \geq {1\over 2}$, it is sufficient that 
\[
C \geq {2\lfloor\log_2 x_l\rfloor + 1 \over \log_2(1+{x_l\over 2})}.
\]
The function of the right hand side takes maximum when $x_l=4$ and corresponding $C$ value is $5/\log_2 3$ which is $\leq 3.1547$. 

We can get improved bound if we do the analysis 2 segments at a a time; that is, inductively assume the bound for $\sum_{i=1}^{l-2} G(x_i)$ and expand the last two segments. This will lead to an improved bound of $2.7024\log_2 n$ (the exact irrational being $10/\log_2 13$). If we do a more complicated analysis taking 3 segments at a time, we will get the bound $2.6646\log_2 n$. Here we give an analysis taking 4 segments at a time leading to an upper bound of ${16/\log_2 73}\log_2 n \leq 2.5849\log_2 n$.

\paragraph*{${2.5849\log n}$ Upper Bound}
%Here we will prove an upper bound of $2.5849\log_2n+1$. This 2.5849 is an approximation of the exact bound: $\frac{16}{\log_273}$.
Now we will do an analysis taking four segments at a time to find a better upper bound. The constant $C$ comes out in the analysis as in the earlier case, but we will fix it beforehand to  $C = 2.5849 > {16/\log_2 73}$. 
\[
\sum_{i=1}^lG(x_i) \leq C\log_2d_l.
\]

%If multiple steps are not needed, this section is unnecessary
\begin{comment}
For this proof, we formulate the problem as steps of $k$ or $k+1$ $x_i$. We will consider a step $S$ to be $S_j=\{x_i, x_{i+1},...,x_{i+k}\}$ or $S_j=\{x_i, x_{i+1},...,x_{i+k}, x_{i+k+1}\}$ for some $i$, $j$, and $k$. For any $j_n$ and $j_m$ where $j_n \neq j_m$, then $S_{j_n} \cap S_{j_m} = \varnothing$. The union of all $S_j$ is equal to the series of all $x_i$. We can write any number of $x_i$ $y$ as steps of $k$ and $k+1$ if $y$ is large enough. For $y$ smaller, we can account show individually. This can be seen by first dividing $y$ by $k$ and getting a value $a$. Then we look at the remainder. If the remainder is 0, then we are done. If the remainder is 1, then we remove one from $a$ and now the remainder is $k+1$, divisible by $k+1$. If the remainder is 2, then we remove 2 from $a$ and now the remainder is $2k+2$, divisible by $k+1$. Thus, we must be able to subtract $k$ a maximum of $k-1$ times. Thus, $y$ must be greater than $k(k-1)$. We will do this prove with $k=3$, but this can be done with arbitrarily large $k$, though the task becomes exponentially more difficult.

{\color{red} We need to write this more formally, splitting into k and k+1}. 

We can rewrite the sum of $x_i$ as:
\[
\sum_{j=1}^lG(S_p)=2.6646\log_2d_p+1.
\]
\end{comment}

\noindent We will prove this with induction. The arguments are similar to the warm-up case. However, since we have functions of 4 variables to deal with the analysis is somewhat messy and involves computational verification. 

\paragraph*{Base cases}
%All $y$ where $3a+4b=y$ does not provide an integer solution. This is when $y=1,2,5$.
Since we will be analyzing four segments at a time, we must consider the base cases for when there are four or less total segments $l$ required to find the final rational. 

$l=1$: we must show:
\[
G(x_1) \leq C \log_2d_1.
\]
$d_1 = x_1+1$: it can be verified that for the chosen $C$ and $x_1\geq 1$. 
\[
2\lfloor\log_2x_1\rfloor +1 \leq C\log_2(x_1+1)  
\]

%Since $G(x_1) =2\lfloor\log_2x_1\rfloor +1 \leq 2\log_2x_1+1$, it is sufficient to show:
%\[
%2\log_2x_1+1 \leq C \log_2d_1.
%\]
%We can also write out $d_1=d_0+x_1m_0=1+x_1$:
%\[
%2\log_2x_1+1 \leq C \log_2(1+x_1).
%\]
%We can cancel out the constant term and get the following:
%\[
%2\log_2x_1 +1 \leq C \log_2(1+x_1).
%\]
%To simplify the analysis, we can write $\log_2(x_1)$ for $\log_2(1+x_1)$:
%\[
%2\log_2x_1 +1 \leq C \log_2(x_1).
%\]
%Solving for $x_1$ gives us:
%\[
%x_1 \geq 2^{\frac{1}{C-2}}.
%\]
%Thus, all $x$ that do not satisfy this inequality must be handled separately. When $C=2.5849$, the inequality is satisfied when $x\geq3.4194$. We must handle now the subcases of $x_1=1,2$. Now we go back to the original statement we are proving:
%\[
%G(x_1) \leq C \log_2d_1.
%\]
%We can plug in the $x_1=1,2$ to get the following inequalities
%\[
%x_1=1: 1 \leq 2C,
%\]
%\[
%x_1 = 2: 3 \leq 3C,
%\]
%\[
%x_1 = 3: 3 \leq 4C.
%\]
%Since $C>1$, the inequality holds true in these subcases.

\noindent $l=2$: we should show the following:
\[
G(x_1)+G(x_2) \leq C \log_2d_2
\]
$d_2 = x_1x_2+x_1+1$. We can remove the floor functions in \(G(x_1)\) and \(G(x_2)\). It is sufficient to show
\[
2\log_2x_1x_2+2 \leq C \log_2d_2.
\]
Writing out $d_2$ explicitly in the expression, we get:
\[
2\log_2x_1x_2+2 \leq C \log_2(x_1x_2+x_1+1).
\]
For larger values of $x$s, it is sufficient to show $2\log_2x_1x_2+2 \leq C \log_2(x_1x_2)$. In particular, for 
$x_1x_2 \geq 2^{\frac{2}{C-2}}$, the inequality holds. In particular, plugging in $C=2.5849$, we get that when $x_1x_2 \geq 11$, the inequality holds. The case when $x_1x_2 < 11$, we can explicitly test for all $\{x_1,x_2\mid x_1,x_2 \geq 1 \mbox{ and }x_1x_2 < 11\}$ in the original inequality. 
For when the product does not satisfy this condition, we can plug in the finite number of possible values of $x_1$ and $x_2$ that have a product of less than or equal to 11 into the original inequality. Since we will be using this method repeatedly, we wrote an exhaustive search program to automate this process. In particular the program takes a top value TOP, number of variables $n$, and two functions $f$ and $g$ on $n$ variables and checks that $f \leq g$ for all tuples $\{\langle x_1,\ldots, x_n\rangle \mid x_i \geq 1 \mbox{ and } \prod_i x_i \leq {\rm TOP}\}$. We verified this for the above case for two variables.

$l=3$: We need to show the following:
\[
G(x_1)+G(x_2)+G(x_3) \leq C \log_2d_3
\]
$d_3=x_1x_2x_3+x_1x_2+x_1+x_3+1$.

Once again, for larger values of $x_i$s, it is sufficient to show:
$2\log_2(x_1x_2x_3)+3 \leq C\log_2(x_1x_2x_3)$. For the chosen value of $C$, the inequality holds 
$x_1x_2x_3 \geq 35$. Again, for the smaller cases, the program verified that all the tuples $\{x_1,x_2,x_3\mid x_1,x_2,x_3\geq 1 \mbox{ and }x_1x_2x_3 < 35\}$ satisfies the original inequality $G(x_1)+G(x_2)+G(x_3) \leq C \log_2{d_3}$. 

%Thus, when the product is greater than 34, the statement is true. For the cases when the product is not greater than 34, we must do a similar method to $y=2$.

%Fo $y=3$, we used the aforementioned program and inputted the top range as 34 and number of variables as 3. The program ran without terminating so thus the inequality holds true for that range. This base case is now shown to hold true.

$l=4$: We need to show the following:
\[
G(x_1)+G(x_2)+G(x_3)+G(x_4)\leq C \log_2d_4
\]
Again, using the same method, we find that the simplified inequality holds true when $x_1x_2x_3x_4 \geq 115$ for the chosen $C=2.5849$. For the case when $x_1x_2x_3x_4 < 115$, we computationally verified.  

%We used the same program with a top range of 114 and number of variables 4. The program ran without terminating so thus, the inequality holds for range $x_1x_2x_3x_4 \leq 114$ and this base case is shown to be true.

% not necessary if doing only steps of 4
\begin{comment}
Finally, we can do the same method for $y=5$ for the following inequality:
\[
G(x_1)+G(x_2)+G(x_3)+G(x_4)+G(x_5) \leq 2.6646 \log_2d_5+1
\]
Using the same method, we find that the inequality holds true when $x_1x_2x_3x_4x_5 \geq 64.83$. We inputted to the same program with a top range of 64 and 5 variables. The program ran without premature termination so thus the inequality holds true in the range $x_1x_2x_3x_4x_5 \leq 64$
\end{comment}

\paragraph*{Inductive Step}
For our inductive hypothesis, we assume the following holds true for any integer $p<l$:
\[
\sum_{i=1}^pG(x_i) \leq C \log_2d_p.
\]
%\subsubsection{Inductive Step}
We want to show the following:
\[
\sum_{i=1}^{l}G(x_i) \leq C\log_2d_{l}.
\]
We can rewrite this as
$\sum_{i=1}^{l-4}G(x_i) + G(x_{l-3})+G(x_{l-2}) + G(x_{l-1}) + G(x_{l}) \leq C\log_2d_{l}.$ Assuming the hypothesis, it is sufficient to show:
\[
C\log_2d_{l-4} + G(x_{l-3}) +G(x_{l-2}) + G(x_{l-1}) + G(x_{l}) \leq C\log_2d_{l}.
\]
We can rewrite this as follows:
\begin{align}
\label{eq:equation_to_prove}
G(x_{l-3}) +G(x_{l-2}) + G(x_{l-1}) + G(x_{l}) \leq C\log_2 \left(\frac{d_{l}}{d_{l-4}} \right).
\end{align}

\noindent We can expand out $d_l$ up to four steps (using identities given in Claim~\ref{claim:d})  to be:
\begin{align}
d_l = 
&\left( d_{l-4} + x_{l-3}m_{l-4} \right)(1+x_{l-1}+x_lx_{l-1}) \\ \nonumber
& + \left(d_{l-4}+(x_{l-3}-1)m_{l-4} \right)(x_{l-2}+x_{l-1}x_{l-2}+x_lx_{l-1}x_{l-2}-x_{l-1} - x_lx_{l-1} + x_l)
\end{align}

\noindent We can then write out $d_l / d_{l-4}$ as:
\begin{align}
\frac{d_l}{d_{l-4}} = 
&\left( 1 + x_{l-3} \frac{m_{l-4}}{d_{l-4}} \right)(1+x_{l-1}+x_lx_{l-1}) \\ \nonumber
& + \left(1+(x_{l-3}-1)\frac{m_{l-4}}{d_{l-4}} \right)(x_{l-2}+x_{l-1}x_{l-2}+x_lx_{l-1}x_{l-2}-x_{l-1} - x_lx_{l-1} + x_l)
\end{align}

\noindent Since $\frac{m_l}{d_l} \geq\frac{1}{2}$, it is sufficient to write $d_l/d_{l-4}$ as the following:
\begin{align}
\label{eq:final_dl_dl-4}
\frac{d_l}{d_{l-4}} = 
&\left( 1 +\frac{x_{l-3}}{2} \right)(1+x_{l-1}+x_lx_{l-1}) \\ \nonumber
& + \left(1+\frac{x_{l-3}-1}{2} \right)(x_{l-2}+x_{l-1}x_{l-2}+x_lx_{l-1}x_{l-2}-x_{l-1} - x_lx_{l-1} + x_l)
\end{align}

\noindent From here we can substitute Equation~\ref{eq:final_dl_dl-4} for $d_l/d_{l-4}$ in Equation~\ref{eq:equation_to_prove}. With some further algebraic manipulation, we get that it is sufficient to show the following:

\begin{align}
\label{eq:four_steps}
G(x_{l-3})+G(x_{l-2}) + G(x_{l-1}) + G(x_l) 
&\leq C \log_2 \Bigg( 1 
  + \frac{x_l}{2} 
  + \frac{x_{l-1}}{2} 
  + \frac{x_{l-2}}{2} 
  + \frac{x_{l-3}}{2} \\ \nonumber
&\quad + \frac{x_l x_{l-1}}{2} 
  + \frac{x_{l-1} x_{l-2}}{2} 
  + \frac{x_{l-2} x_{l-3}}{2}  \\ \nonumber
&\quad + \frac{x_l x_{l-3}}{2}
  + \frac{x_l x_{l-1} x_{l-2}}{2} 
  + \frac{x_{l-1} x_{l-2} x_{l-3}}{2} \\ \nonumber
&\quad + \frac{x_l x_{l-1} x_{l-2} x_{l-3}}{2} 
\Bigg) 
\end{align}
where the left hand side is  $2\left(\lfloor \log_2 x_{l-3}\rfloor + \lfloor \log_2 x_{l-2}\rfloor + \lfloor \log_2 x_{l-1}\rfloor + \lfloor \log_2 x_{l-l}\rfloor \right) + 4.$ 

At this point we explain how the constant $C$ is chosen. Let $f(x_{l-3},x_{l-2},x_{l-1},x_{l})$ be the LHS function and $g(x_{l-3},x_{l-2},x_{l-1},x_{l})$ be the RHS function. Then we want $C\geq f/g$. By inspecting the function, we can conclude that   for larger values of $x_i$s, the inequality should hold for $C>2$. Since the problem case is when $x_i$s are small, we can run an exhaustive search for the worst-case $x_i$s in a bounded region. With sufficient search parameters, we find the values $x_l=4,x_{l-1}=2,x_{l-2}=2,x_{l-3}=4$ to have the maximum value for $f/g$. From Equation~\ref{eq:four_steps}, we can plug in these values to get the following:
\[
5+3+3+5 \leq C \log_2\left( 1+ \frac{4}{2} + \frac{2}{2} + \frac{2}{2} + \frac{4}{2}+ \frac{8}{2} + \frac{4}{2} + \frac{8}{2} + \frac{16}{2} + \frac{16}{2} + \frac{16}{2} + \frac{64}{2}\right) 
\]
Solving for $C$ gives us the constant we use:
\[
C = \frac{16}{\log_273} \approx 2.5849.
\]
We need to show the inequality holds with this chosen $C$. Since $G(x) = 2\lfloor \log_2 x\rfloor + 2 \leq 2\log_2 x + 2$, for larger values of $x_i$s, we can show that even 
\[
2\log_2(x_{l-3}x_{l-2}x_{l-1}x_l)+4 \leq C \log_2\left( \frac{x_{l}x_{l-1}x_{l-2}x_{l-3}}{2} \right).
\]
In particular, this holds for 
$
x_{l-3}x_{l-2}x_{l-1}x_l \geq 2^{\frac{4+C}{C-2}}.$
For $C=2.5849$, the inequality holds true when $x_{l-3}x_{l-2}x_{l-1}x_l \geq 2450$. For the case when $x_{l-3}x_{l-2}x_{l-1}x_l < 2450$, there is only a finite number of tuples to test and we can use the aforementioned program to verify that the
inequality given in Equation~\ref{eq:four_steps} holds for all possible combinations of such $x_i$.

\begin{comment}
For the case that some $x_i \geq 64$, a different method must be used to show the inequality holds true. If at least one is greater than 64, then the following equation would be true:

plug in $x_i$ for when one is 64 and the rest are 1.
\end{comment}

\smallskip
\noindent{\em Remark:} 
We can employ the same method as above with larger values of segments (with five for example) and we suspect that the bound can be brought closer to our lower bound of $2.4189\log_2 n$. However, that is uninspiring and we do not know how to do a cleaner analysis to get the exact bound.

\subsection{$2.4189\log_2 n$ Lower Bound}

Here we prove a lower bound of approximately $2.4189\log_2n$ for the wort-case number of comparisons required by the algorithm. We will show this by finding the rate of increase of the denominators for  a series of alternating left and right traversals in the Stern-Brocot tree. For two integers $a$ and $b$, consider rationals represented by $(L^aR^b)^k$ for increasing values of $k$. Notice that the number of comparisons used by the algorithm for these rationals is $k\times(G(a)+G(b))$. We will express $k$ as a function of $n$ for large values of $n$. 

Recall Claim~\ref{claim:d}: for any $i$, $d_i=d_{i-1} + x_im_{i-1}$, $m_i = d_{i-1} + (x_i-1)m_{i-1}$, and $d_i = m_i + m_{i-1}$. From this we have:
%We can rewrite $m_{k-1}$ as $d_{k-1}-m_{k-2}$ to give us $d_k=d_{k-1} + x_k(d_{k-1}-m_{k-2})$. $m_{k-2}$ can be written as $\frac{d_{k-1}-d_{k-2}}{x_{k-1}}$ to give us the following:
\[
d_i=d_{i-1} + x_i \left( d_{i-1}-\frac{d_{i-1}-d_{i-2}}{x_{i-1}} \right).
\]
Dividing by $d_{i-1}$ we get:
\[
\frac{d_i}{d_{i-1}}=1 + x_i - \frac{x_i}{x_{i-1}} + \frac{d_{i-2}}{d_{i-1}} \cdot \frac{x_i}{x_{i-1}}.
\]
Similarly, the ratio between $d_{i-1}$ and $d_{i-2}$ is:
\[
\frac{d_{i-1}}{d_{i-2}}=1 + x_{i-1} - \frac{x_{i-1}}{x_{i-2}} + \frac{d_{i-3}}{d_{i-2}} \cdot \frac{x_{i-1}}{x_{i-2}}.
\]

Assume that all $x_i$ are alternating numbers $a,b,a,b,\ldots$. Let $\phi_a$ denote the ratio $d_i/d_{i-1}$ in the limit for odd $i$s and $\phi_b$ denote the ratio after $b$ traversals in the limit  for even $i$s. 
Then have the following system of equations:
\[
\phi_a=1 + a - \frac{a}{b} + \phi_b^{-1} \cdot \frac{a}{b}
\mbox{ and }
\phi_b=1 + b - \frac{b}{a} + \phi_a^{-1} \cdot \frac{b}{a}.
\]
%{\color{red} Should I add more details on solving this equation}
Solving this system of equations for $\phi_a$ and $\phi_b$ will give:
\[
\phi_a = \frac{2-2\frac{b}{a}+ab+\sqrt{4ab+a^2b^2}}{2(1+b-\frac{b}{a})} \mbox{ and }
\phi_b = \frac{2-2\frac{a}{b}+ab+\sqrt{4ab+a^2b^2}}{2(1+a-\frac{a}{b})}.
\]

From this, $k=\log_{\phi_a\phi_b} n$ and the overall number of comparisons to to reach a denominator at $n$ for $(L^aR^b)^k$ is:
\[
(G(a)+G(b)) \log_{\phi_a\phi_b}n = 
\frac{(G(a)+G(b))}{\log_2\phi_a\phi_b}\log_2n.
\]

To find the worst case $a$ and $b$,  we tested all combinations of $a$s and $b$s ranging from 1 to 1000. The worst case value was the sequences with $a=8$ and $b=1$ which results in $\frac{8}{\log_2(5+2\sqrt{6})}\log_2 n > 2.4189\log_2n$ number of comparisons. Thus searching for fraction of the form $(L^8R^1)^k$ takes at least $2.4189\log_2n$ number of comparisons.

\subsection{Best Rational Approximation for Unknown Reals}
We can modify the algorithm to solve the best rational approximation problem for an unknown real number $r$. 
Recall that the rational approximation problem for an unknown real number $r$ is: given an approximation parameter $\delta$, find the (unique) fraction $a\over b$ so that ${a\over b} \in [r\pm \delta]$ with smallest possible denominator $b$. 

The modified algorithm works in much the same way as the rational search algorithm. When traversing to the left, fractions are queried the same to find some $x$ so that $\frac{p' + xp}{q'+xq} \leq \alpha \leq \frac{p' + (x-1)p}{q'+(x-1)q}$. When traversing to the right, fractions are queried to find $x$ so $\frac{p + xp'}{q+xq'} \geq \alpha \geq \frac{p + (x-1)p'}{q+(x-1)q'}$. This is done with exponential and binary search as in the rational search. Once we find $x$, we want to see if the fractions at $x$ or $x-1$ traversals are in the range $[\alpha-\delta, \alpha + \delta]$. In the case of going left, we know $\frac{p' + xp}{q'+xq} < \alpha$. To see if  $\frac{p' + xp}{q'+xq}$ is within the range, we can query to see if $\frac{p' + xp}{q'+xq} + \delta \geq \alpha$. For the fraction at $x-1$ traversals to the left, we can query to see if $\frac{p' + (x-1)p}{q'+(x-1)q} - \delta \leq \alpha$. When traversing to the right, similar queries can be called. If the fractions at $x$ and $x-1$ traversals are not within the range, then we continue our search but switch to the opposite direction in the same manner as rational search. If the fraction at $x$ traversals is within the range and not the fraction at $x-1$ traversals, then we return the fraction at $x$ traversals. If the fraction at $x-1$ traversals is in range, then a binary search must be done from $1$ to $x-1$ traversals to find the number of traversals $z$. The fraction at $z$ traversals will be in the range $[\alpha\pm \delta]$ and the fraction at $z-1$ traversals will not. We return the fraction at $z$ traversals. 

\smallskip
\noindent{\em Remark:} 
While the algorithm and its complexity analysis are formulated for rational or real numbers in the interval \([0,1]\), it can be naturally extended to operate over the generalized Stern--Brocot tree, which spans all positive rationals in the interval \((0, \infty)\). In fact, for our experiments on comparison-based rational approximation of real numbers, we implemented this generalized version of the algorithm.

\subsection{Kwek-Mehlhorn Algorithm}
\label{kwek-mehlhorn}
We briefly describe Kwek-Mehlhorn rational search algorithm here. Given an an unknown rational $\alpha$,   
%To solve the rational search problem, Kwek and Mehlhorn proposed an algorithm~\cite{kwek2003optimal} that takes \(2\log_2 n + O(1) \) queries with an additional \(2\log_2n \) processing time to find any rational \(x=\frac{a}{b} \) with \(a \leq n\) and \(b \leq n\). They define the set of all such rationals as \(Q_n\). 
The algorithm has two phases. The first phase finds a $\mu$ such that \( \alpha \in [\frac{\mu}{n^2},\frac{\mu+1}{n^2}]\) using a standard binary search with comparison queries (which takes \(2\log_2n\) queries). The range between the two fractions \( \frac{\mu}{n^2} \) and \( \frac{\mu+1}{n^2}\) is \(\frac{1}{n^2}\). They observe that for any two distinct  fractions \((\frac{a}{b},\frac{c}{d}) \) such that $b,d \leq n$, the following holds:
\(
\left| \frac{a}{b} - \frac{c}{d}\right| = \left| \frac{ac-bd}{bd} \right| \geq \frac{1}{n^2}.
\)
From this, we can see that there must be only one fraction \( \frac{a}{b}\) so that $b\leq n$ in the interval. In the second phase, they identify the unique fraction $a\over b$ with $b\leq n$ in the range \([\frac{\mu}{n^2},\frac{\mu+1}{n^2}] \). For this they use the continued fraction algorithm to find the fraction with the smallest denominator in the range using Euclid's GCD algorithm. Note that, as described, knowing the bound $n$ on the denominators is critical for KM algorithm and it is an interesting open question whether we can design close to  $2\log_2 n + O(1)$ query algorithm for unbounded rational search.

%% file: experiments-writeup.tex
\section{Experiments}

%We ran experiments on the new algorithm and the Kwek-Mehlhorn algorithm. These experiments were set up by running 1000 iterations on each value of $n$ which were powers of 10. At each iteration, a random integer $b$ such that $2\leq b \leq n$. A random integer $a$ such that $1\leq b$ was also generated. The rational $\frac{a}{b}$ and integer $n$ were passed as the input to the Kwek-Mehlhorn algorithm and the Compressed Stern-Brocot Search. The outputted \(\frac{a}{b} \) for each algorithm was checked with the input to verify its correctness. The number of queries were recorded at each iteration with the average time, average queries, and maximum queries across all iterations of $n$ being reported. Tables \ref{tab:rat_search_metrics1} and \ref{tab:rat_search_metrics2} contain the data from these experiments. Figure \ref{fig:rat_search_queries} displays the query results graphically. When computing with such high $n$ and dealing with rationals, Python's native float precision would not yield accurate results. Python's \texttt{decimal} module was used to increase the precision of the floats to ensure the correct result. The Python code will be made available once the paper becomes public. 

We conducted experiments comparing the performance of the new algorithm with the Kwek--Mehlhorn algorithm. Each experiment consisted of 1000 iterations for each value of \( n \), where \( n \) ranged over powers of 10. In each iteration, a random integer \( b \) was chosen such that \( 2 \leq b \leq n \), followed by a random integer \( a \) such that \( 1 \leq a < b \). The fraction \( \frac{a}{b} \) and the corresponding value of \( n \) were used as inputs to both the Kwek--Mehlhorn algorithm and the Compressed Stern--Brocot Search algorithm.
The output  returned by each algorithm was checked against the input to verify correctness. For each iteration, the number of queries made was recorded. We report the average time, average number of queries, and maximum number of queries across all iterations for each value of \( n \).

Tables~\ref{tab:rat_search_metrics1} and~\ref{tab:rat_search_metrics2} summarize the experimental results, and Figure~\ref{fig:rat_search_queries} presents the query statistics graphically. Due to the high precision required for computations involving large \( n \) and rational values, Python's native floating-point arithmetic was insufficient. Therefore, we used Python’s \texttt{decimal} module to ensure adequate numerical precision. The source code will be available upon request. 

The rational search algorithm was also modified and implemented to solve the rational approximation problem for any unknown real number using comparison queries. We used Python's \texttt{math} module for the irrational numbers $\pi$, $e$, $\sqrt{2}$, and $\sqrt{5}$. For every $i\in[1,15]$, the approximation for each irrational number was calculated with precision $10^{-i}$. This solution was verified with our implementation of the Fori{\v{s}}ek approximation algorithm. See Table~\ref{tab:rat_approx_values} for the values calculated. Each approximation was done 1000 times and the average time to calculate the solution is reported. The average time can be found in Table~\ref{tab:rat_approx_time} along with the number of queries used to find the approximation.

All experiments were conducted on a cluster at the resident computing center. Each experiment was submitted as an independent job to a single node. Each node is equipped with two Intel Xeon Gold 6348 CPUs, providing a total of 56 cores.
Memory allocation for each job was adjusted based on preliminary test results to ensure sufficient resources for subsequent runs. 

%All experiments were ran using the Swan cluster in the resident computing center. Every experiment was submitted as an independent job to a single node. The memory allocated in each experiment was modified based on prior test results to ensure that the subsequent experiments had ample memory to use. Each node contains 2 Intel Xeon Gold 6348 CPUs and 56 cores.

When comparing the Kwek-Mehlhorn and Compressed Stern-Brocot Tree Search on the rational search problem, the Kwek-Mehlhorn algorithm scales better when $n$ is  really large. This aligns with the theoretical upper bounds on query complexity,
 although the Kwek-Mehlhorn algorithm does not outperform on average queries in our test distribution until $n \geq 10^{41}$. The average time for the Compressed Stern-Brocot Tree Search is consistently higher even when there is a lower number of average queries. This suggests improvements to the implementation are possible to lower the average time per query. However, in situations where implementation of a query takes substantial time, the savings in number of queries of typical inputs by the new algorithm could be of advantage. The maximum number of queries for the Kwek-Mehlhorn is the same as the average number of queries. Contrasting this, the Compressed Stern-Brocot Tree Search has a wide range between its average queries and maximum queries. Even when the average queries is lower than that of the Kwek-Mehlhorn algorithm, the maximum queries is significantly higher. Though the maximum queries for the Compressed Stern-Brocot Search is much higher, it is lower than the upper bound of \(2.5849 \log_2n\) that we establish. 

%% file: experiments.tex
\begin{table}[]
    \centering
    \renewcommand{\arraystretch}{1.2}
    \begin{tabular}{|c||ccc|ccc|}
        \hline
        \multirow{2}{*}{\(n\)} & \multicolumn{3}{c|}{Kwek-Mehlhorn} & \multicolumn{3}{c|}{Compressed Stern-Brocot} \\ \cline{2-7}
        & Max Queries & Average & Time (s) & Max Queries & Average & Time (s) \\ \hline
        \( 10^1 \) & 8 & 5.5 & \(1.52 \times 10^{-5}\) & 7 & 3.5 & \(2.29 \times 10^{-5}\) \\ \hline
        \( 10^2 \) & 15 & 13.2 & \(2.80 \times 10^{-5}\) & 15 & 8.5 & \(4.42 \times 10^{-5}\) \\ \hline
        \( 10^3 \) & 21 & 20.7 & \(4.19 \times 10^{-5}\) & 23 & 15.1 & \(6.93 \times 10^{-5}\) \\ \hline
        \( 10^4 \) & 28 & 27.6 & \(5.63 \times 10^{-5}\) & 29 & 21.8 & \(9.57 \times 10^{-5}\) \\ \hline
        \( 10^5 \) & 35 & 34.3 & \(6.74 \times 10^{-5}\) & 38 & 28.5 & \(1.17 \times 10^{-4}\) \\ \hline
        \( 10^6 \) & 41 & 40.9 & \(8.11 \times 10^{-5}\) & 44 & 35.5 & \(1.43 \times 10^{-4}\) \\ \hline
        \( 10^7 \) & 48 & 47.6 & \(9.41 \times 10^{-5}\) & 52 & 42.1 & \(1.67 \times 10^{-4}\) \\ \hline
        \( 10^8 \) & 55 & 54.2 & \(1.08 \times 10^{-4}\) & 59 & 48.9 & \(1.92 \times 10^{-4}\) \\ \hline
        \( 10^9 \) & 61 & 60.8 & \(1.22 \times 10^{-4}\) & 66 & 55.8 & \(2.17 \times 10^{-4}\) \\ \hline
        \( 10^{10} \) & 68 & 67.5 & \(2.71 \times 10^{-4}\) & 72 & 62.3 & \(3.78 \times 10^{-4}\) \\ \hline
        \( 10^{11} \) & 75 & 74.1 & \(3.16 \times 10^{-4}\) & 80 & 69.2 & \(4.33 \times 10^{-4}\) \\ \hline
        \( 10^{12} \) & 81 & 80.8 & \(3.71 \times 10^{-4}\) & 86 & 76.2 & \(5.01 \times 10^{-4}\) \\ \hline
        \( 10^{13} \) & 88 & 87.4 & \(4.21 \times 10^{-4}\) & 96 & 82.8 & \(5.61 \times 10^{-4}\) \\ \hline
        \( 10^{14} \) & 95 & 94.0 & \(4.65 \times 10^{-4}\) & 102 & 89.9 & \(6.15 \times 10^{-4}\) \\ \hline
        \( 10^{15} \) & 101 & 100.8 & \(5.22 \times 10^{-4}\) & 110 & 96.4 & \(6.85 \times 10^{-4}\) \\ \hline
        \( 10^{16} \) & 108 & 107.4 & \(5.75 \times 10^{-4}\) & 114 & 103.3 & \(7.51 \times 10^{-4}\) \\ \hline
        \( 10^{17} \) & 114 & 114.0 & \(6.33 \times 10^{-4}\) & 122 & 110.3 & \(8.21 \times 10^{-4}\) \\ \hline
        \( 10^{18} \) & 121 & 120.6 & \(6.85 \times 10^{-4}\) & 128 & 117.0 & \(8.95 \times 10^{-4}\) \\ \hline
        \( 10^{19} \) & 128 & 127.3 & \(7.70 \times 10^{-4}\) & 136 & 123.6 & \(9.81 \times 10^{-4}\) \\ \hline
        \( 10^{20} \) & 134 & 133.9 & \(8.21 \times 10^{-4}\) & 145 & 130.4 & \(1.05 \times 10^{-3}\) \\ \hline
        \( 10^{21} \) & 141 & 140.6 & \(8.73 \times 10^{-4}\) & 151 & 137.4 & \(1.13 \times 10^{-3}\) \\ \hline
        \( 10^{22} \) & 148 & 147.2 & \(9.20 \times 10^{-4}\) & 157 & 144.1 & \(1.21 \times 10^{-3}\) \\ \hline
        \( 10^{23} \) & 154 & 153.9 & \(9.71 \times 10^{-4}\) & 165 & 150.6 & \(1.28 \times 10^{-3}\) \\ \hline
        \( 10^{24} \) & 161 & 160.6 & \(1.02 \times 10^{-3}\) & 172 & 157.6 & \(1.36 \times 10^{-3}\) \\ \hline
        \( 10^{25} \) & 168 & 167.1 & \(1.08 \times 10^{-3}\) & 176 & 164.3 & \(1.45 \times 10^{-3}\) \\ \hline
    \end{tabular}
    \smallskip
    \caption{\centering The max queries, average queries, and average time of the Kwek-Mehlhorn and Compressed Stern-Brocot Search algorithm for performing a rational search on a fraction \(\frac{a}{b} \) such that \(a<b \leq n\), for values of $n\leq 10^{25}$.}
    \label{tab:rat_search_metrics1}
\end{table}

\begin{table}[]
    \centering
    \renewcommand{\arraystretch}{1.2}
    \begin{tabular}{|c||ccc|ccc|}
        \hline
        \multirow{2}{*}{\(n\)} & \multicolumn{3}{c|}{Kwek-Mehlhorn} & \multicolumn{3}{c|}{Compressed Stern-Brocot} \\ \cline{2-7}
        & Max Queries & Average & Time (s) & Max Queries & Average & Time (s) \\ \hline
        \( 10^{26} \) & 174 & 173.8 & \(1.12 \times 10^{-3}\) & 185 & 171.3 & \(1.52 \times 10^{-3}\) \\ \hline
        \( 10^{27} \) & 181 & 180.5 & \(1.17 \times 10^{-3}\) & 195 & 178.2 & \(1.59 \times 10^{-3}\) \\ \hline
        \( 10^{28} \) & 188 & 187.0 & \(1.22 \times 10^{-3}\) & 200 & 184.7 & \(1.67 \times 10^{-3}\) \\ \hline
        \( 10^{29} \) & 194 & 193.7 & \(1.34 \times 10^{-3}\) & 206 & 191.6 & \(1.82 \times 10^{-3}\) \\ \hline
        \( 10^{30} \) & 201 & 200.4 & \(1.40 \times 10^{-3}\) & 213 & 198.4 & \(1.90 \times 10^{-3}\) \\ \hline
        \( 10^{31} \) & 207 & 207.0 & \(1.46 \times 10^{-3}\) & 219 & 205.3 & \(1.99 \times 10^{-3}\) \\ \hline
        \( 10^{32} \) & 214 & 213.7 & \(1.53 \times 10^{-3}\) & 228 & 211.8 & \(2.09 \times 10^{-3}\) \\ \hline
        \( 10^{33} \) & 221 & 220.3 & \(1.59 \times 10^{-3}\) & 235 & 218.4 & \(2.18 \times 10^{-3}\) \\ \hline
        \( 10^{34} \) & 227 & 226.9 & \(1.64 \times 10^{-3}\) & 239 & 225.9 & \(2.25 \times 10^{-3}\) \\ \hline
        \( 10^{35} \) & 234 & 233.6 & \(1.71 \times 10^{-3}\) & 247 & 232.5 & \(2.35 \times 10^{-3}\) \\ \hline
        \( 10^{36} \) & 241 & 240.2 & \(1.77 \times 10^{-3}\) & 253 & 239.0 & \(2.44 \times 10^{-3}\) \\ \hline
        \( 10^{37} \) & 247 & 246.9 & \(1.83 \times 10^{-3}\) & 264 & 245.7 & \(2.54 \times 10^{-3}\) \\ \hline
        \( 10^{38} \) & 254 & 253.6 & \(1.99 \times 10^{-3}\) & 268 & 252.9 & \(2.72 \times 10^{-3}\) \\ \hline
        \( 10^{39} \) & 261 & 260.1 & \(2.07 \times 10^{-3}\) & 274 & 259.7 & \(2.83 \times 10^{-3}\) \\ \hline
        \( 10^{40} \) & 267 & 266.8 & \(2.13 \times 10^{-3}\) & 282 & 265.9 & \(2.93 \times 10^{-3}\) \\ \hline
        \( 10^{41} \) & 274 & 273.5 & \(2.22 \times 10^{-3}\) & 289 & 273.2 & \(3.05 \times 10^{-3}\) \\ \hline
        \( 10^{42} \) & 281 & 280.0 & \(2.30 \times 10^{-3}\) & 298 & 280.2 & \(3.17 \times 10^{-3}\) \\ \hline
        \( 10^{43} \) & 287 & 286.7 & \(2.34 \times 10^{-3}\) & 303 & 286.7 & \(3.23 \times 10^{-3}\) \\ \hline
        \( 10^{44} \) & 294 & 293.4 & \(2.41 \times 10^{-3}\) & 311 & 293.8 & \(3.34 \times 10^{-3}\) \\ \hline
        \( 10^{45} \) & 300 & 300.0 & \(2.50 \times 10^{-3}\) & 316 & 300.5 & \(3.47 \times 10^{-3}\) \\ \hline
        \( 10^{46} \) & 307 & 306.7 & \(2.57 \times 10^{-3}\) & 323 & 307.1 & \(3.57 \times 10^{-3}\) \\ \hline
        \( 10^{47} \) & 314 & 313.3 & \(2.64 \times 10^{-3}\) & 335 & 313.7 & \(3.67 \times 10^{-3}\) \\ \hline
        \( 10^{48} \) & 320 & 319.9 & \(2.83 \times 10^{-3}\) & 339 & 321.2 & \(3.90 \times 10^{-3}\) \\ \hline
        \( 10^{49} \) & 327 & 326.6 & \(2.90 \times 10^{-3}\) & 345 & 327.6 & \(4.00 \times 10^{-3}\) \\ \hline
        \( 10^{50} \) & 334 & 333.3 & \(2.97 \times 10^{-3}\) & 351 & 334.0 & \(4.10 \times 10^{-3}\) \\ \hline
    \end{tabular}
    \smallskip
    \caption{\centering The max queries, average queries, and average time of the Kwek-Mehlhorn and Compressed Stern-Brocot Search algorithm for performing a rational search on a fraction \(\frac{a}{b} \) such that \(a<b \leq n\), for values of $n>10^{25}$.}
    \label{tab:rat_search_metrics2}
\end{table}

\begin{figure}
    \centering
    \includegraphics[width=\linewidth]{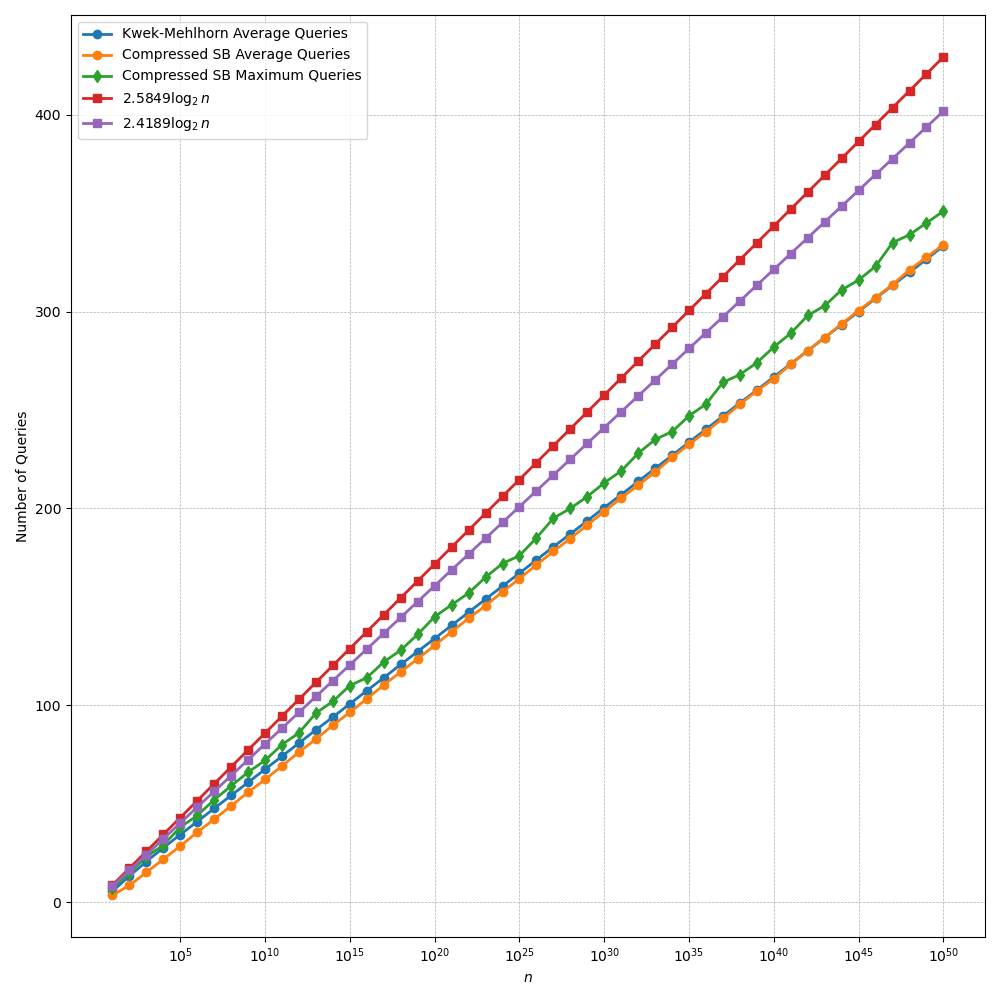}
    \caption{\centering Queries taken by the Kwek-Mehlhorn and Compressed Stern-Brocot Tree Search algorithm for performing a rational search on a fraction \(\frac{a}{b} \) such that \(a<b \leq n\). Note that the maximum queries for the Kwek-Mehlhorn algorithm were not displayed in this graph because they are effectively the same as the average.}
    \label{fig:rat_search_queries}
\end{figure}

\begin{table}[]
    \centering
    \begin{tabular}{|c|c|c|c|c|}
        \hline
        $\delta$ & $\pi$ & $e$ & $\sqrt{2}$ & $\sqrt{5}$ \\ \hline
        $10^{-1}$ & $16/5$ & $8/3$ & $3/2$ & $7/3$ \\ \hline
        $10^{-2}$ & $22/7$ & $19/7$ & $17/12$ & $29/13$ \\ \hline
        $10^{-3}$ & $201/64$ & $87/32$ & $41/29$ & $38/17$ \\ \hline
        $10^{-4}$ & $333/106$ & $193/71$ & $99/70$ & $161/72$ \\ \hline
        $10^{-5}$ & $355/113$ & $1071/394$ & $577/408$ & $682/305$ \\ \hline
        $10^{-6}$ & $355/113$ & $2721/1001$ & $1393/985$ & $2207/987$ \\ \hline
        $10^{-7}$ & $75948/24175$ & $15062/5541$ & $3363/2378$ & $9349/4181$ \\ \hline
        $10^{-8}$ & $100798/32085$ & $23225/8544$ & $19601/13860$ & $12238/5473$ \\ \hline
        $10^{-9}$ & $103993/33102$ & $49171/18089$ & $47321/33461$ & $51841/23184$ \\ \hline
        $10^{-10}$ & $312689/99532$ & $419314/154257$ & $114243/80782$ & $219602/98209$ \\ \hline
        $10^{-11}$ & $833719/265381$ & $1084483/398959$ & $275807/195025$ & $710647/317811$ \\ \hline
        $10^{-12}$ & $4272943/1360120$ & $1084483/398959$ & $1607521/1136689$ & $3010349/1346269$ \\ \hline
        $10^{-13}$ & $5419351/1725033$ & $12496140/4597073$ & $3880899/2744210$ & $3940598/1762289$ \\ \hline
        $10^{-14}$ & $58466453/18610450$ & $28245729/10391023$ & $9369319/6625109$ & $16692641/7465176$ \\ \hline
        $10^{-15}$ & $80143857/25510582$ & $28245729/10391023$ & $54608393/38613965$ & $70711162/31622993$ \\ \hline
    \end{tabular}
    \smallskip
    \caption{\centering Rational approximations returned by our algorithm for given $\delta$ and $\alpha$.}
    \label{tab:rat_approx_values}
\end{table}

\begin{table}[]
    \centering
    \renewcommand{\arraystretch}{1.2}  % Optional: adds a bit of vertical spacing
    \begin{tabular}{|c||cc|cc|cc|cc|}
        \hline
        \multirow{2}{*}{\(\delta\)}
        & \multicolumn{2}{c|}{$\pi$}
        & \multicolumn{2}{c|}{$e$}
        & \multicolumn{2}{c|}{$\sqrt{2}$}
        & \multicolumn{2}{c|}{$\sqrt{5}$} \\
        \cline{2-9}
        & Time(s) & Q & Time(s) & Q & Time(s) & Q & Time(s) & Q\\ \hline
        $10^{-1}$ & $7.036\times 10^{-6}$ & $14$ & $5.527\times 10^{-6}$ & $12$ & $3.889\times 10^{-6}$ & $6$ & $5.662\times 10^{-6}$ & $11$ \\ \hline
        $10^{-2}$ & $5.744\times 10^{-6}$ & $11$ & $6.728\times 10^{-6}$ & $13$ & $7.143\times 10^{-6}$ & $10$ & $7.758\times 10^{-6}$ & $16$ \\ \hline
        $10^{-3}$ & $1.049\times 10^{-5}$ & $24$ & $9.246\times 10^{-6}$ & $17$ & $8.588\times 10^{-6}$ & $13$ & $7.751\times 10^{-6}$ & $16$ \\ \hline
        $10^{-4}$ & $1.037\times 10^{-5}$ & $24$ & $1.252\times 10^{-5}$ & $17$ & $1.024\times 10^{-5}$ & $14$ & $9.539\times 10^{-6}$ & $17$ \\ \hline
        $10^{-5}$ & $9.513\times 10^{-6}$ & $19$ & $1.430\times 10^{-5}$ & $26$ & $1.358\times 10^{-5}$ & $18$ & $1.256\times 10^{-5}$ & $24$ \\ \hline
        $10^{-6}$ & $9.475\times 10^{-6}$ & $19$ & $1.603\times 10^{-5}$ & $27$ & $1.674\times 10^{-5}$ & $21$ & $1.537\times 10^{-5}$ & $27$ \\ \hline
        $10^{-7}$ & $1.978\times 10^{-5}$ & $47$ & $1.991\times 10^{-5}$ & $34$ & $1.732\times 10^{-5}$ & $22$ & $1.785\times 10^{-5}$ & $32$ \\ \hline
        $10^{-8}$ & $2.081\times 10^{-5}$ & $47$ & $1.992\times 10^{-5}$ & $34$ & $2.103\times 10^{-5}$ & $26$ & $1.786\times 10^{-5}$ & $32$ \\ \hline
        $10^{-9}$ & $1.918\times 10^{-5}$ & $47$ & $2.113\times 10^{-5}$ & $33$ & $2.221\times 10^{-5}$ & $29$ & $1.998\times 10^{-5}$ & $33$ \\ \hline
        $10^{-10}$ & $1.891\times 10^{-5}$ & $39$ & $2.591\times 10^{-5}$ & $45$ & $2.548\times 10^{-5}$ & $30$ & $2.313\times 10^{-5}$ & $40$ \\ \hline
        $10^{-11}$ & $2.133\times 10^{-5}$ & $46$ & $2.631\times 10^{-5}$ & $45$ & $2.540\times 10^{-5}$ & $33$ & $2.606\times 10^{-5}$ & $43$ \\ \hline
        $10^{-12}$ & $2.424\times 10^{-5}$ & $50$ & $2.709\times 10^{-5}$ & $45$ & $2.879\times 10^{-5}$ & $37$ & $2.841\times 10^{-5}$ & $48$ \\ \hline
        $10^{-13}$ & $2.360\times 10^{-5}$ & $47$ & $3.199\times 10^{-5}$ & $54$ & $3.234\times 10^{-5}$ & $38$ & $2.829\times 10^{-5}$ & $48$ \\ \hline
        $10^{-14}$ & $2.860\times 10^{-5}$ & $60$ & $3.332\times 10^{-5}$ & $54$ & $3.403\times 10^{-5}$ & $41$ & $2.954\times 10^{-5}$ & $49$ \\ \hline
        $10^{-15}$ & $2.853\times 10^{-5}$ & $60$ & $3.789\times 10^{-5}$ & $63$ & $3.660\times 10^{-5}$ & $45$ & $3.243\times 10^{-5}$ & $56$ \\ \hline
    \end{tabular}
    \smallskip
    \caption{\centering Time and queries (columns abbreviated as Q) taken to compute the rational approximation for given $\delta$ and $\alpha$ by our algorithm.}
    \label{tab:rat_approx_time}
\end{table}